\documentclass[12pt]{article}
\usepackage{epsfig}

\newcommand{\mysection}{\setcounter{equation}{0}\section}

\def\beq{\begin{equation}}
\def\eeq{\end{equation}}
\def\beqa{\begin{eqnarray}}
\def\eeqa{\end{eqnarray}}

\newlength{\dinwidth} \newlength{\dinmargin}
\begin{document}

\begin{center}
{\Large \bf NNLO soft-gluon corrections for the $Z$-boson and
$W$-boson transverse momentum distributions}
\end{center}
\vspace{2mm}
\begin{center}
{\large Nikolaos Kidonakis$^a$ and Richard J. Gonsalves$^b$}\\
\vspace{2mm}
${}^a${\it Kennesaw State University,  Physics \#1202,\\
1000 Chastain Rd., Kennesaw, GA 30144-5591, USA} \\
\vspace{2mm}
${}^b${\it Department of Physics, University at Buffalo,
The State University of New York, \\ Buffalo, NY 14260-1500, USA}
\end{center}

\begin{abstract}
We present results for the $Z$-boson and $W$-boson transverse momentum ($p_T$) distributions for large $p_T$ at LHC and Tevatron energies. We calculate complete next-to-leading-order (NLO) QCD corrections as well as soft-gluon corrections at next-to-next-to-leading-order (NNLO) to the differential cross section.
The NNLO soft-gluon contributions are derived from next-to-next-to-leading-logarithm (NNLL) resummation at two loops.
We find enhancements of the $p_T$ distributions and reductions of the scale dependence when the NNLO corrections are included.
\end{abstract}

\mysection{Introduction}

The production of $Z$ and $W$ bosons with large transverse momentum, $p_T$, has been observed and analyzed at both the Tevatron \cite{CDF,D0} and the
LHC \cite{ATLAS,CMS}. The study of electroweak boson production complements studies of Higgs physics and top quark production in the Standard Model.
Furthermore, these processes are backgrounds to new physics that may be within reach of the LHC, and thus it is important to have precise theoretical predictions.
High-$p_T$ $W$ production has a clean experimental signature when an on-shell
$W$ decays to leptons, and accurate predictions are needed to
reduce uncertainties in precision measurements of the $W$ mass and decay width.
The charged leptons in complementary processes involving $Z$ bosons
can be measured with somewhat higher resolution than the neutrino,
but the observed event rate for $W$ bosons at the LHC is larger than that for $Z$ bosons produced on shell.  The event rate for
the underlying production mechanism is enhanced experimentally by measuring
off-shell $Z$ bosons and virtual photons decaying to lepton pairs.

At leading order (LO) in the strong coupling, $\alpha_s$, an electroweak boson
can be produced with large $p_T$ by recoiling
against a single parton which decays into a jet of hadrons.
The LO partonic processes for $Z$ production at large $p_T$ are
$q g \rightarrow Z q$ and $q {\bar q} \rightarrow Z g$, and for $W$ production
they are $q g \rightarrow W q'$ and $q {\bar q'} \rightarrow W g$.
The next-to-leading-order (NLO) corrections, involving virtual one-loop graphs and two-parton final states, for $Z$ and $W$ production at large $p_T$ were calculated in \cite{AR,gpw} where complete analytic expressions were provided.
The NLO corrections enhance the differential $p_T$ distributions and they
reduce the factorization and renormalization scale dependence.

Higher-order contributions to electroweak-boson production from
the emission of soft gluons have also been calculated.
These corrections appear in the form of logarithms which can be formally resummed, and they were first calculated to
next-to-leading-logarithm (NLL) accuracy in \cite{NKVD} using the moment-space resummation formalism in perturbative Quantum ChromoDynamics (pQCD).
Approximate next-to-next-to-leading-order (NNLO) corrections derived from the resummation were calculated in \cite{NKASV} and were shown to provide enhancements and a further reduction of the scale dependence. Numerical results for the $W$-boson $p_T$ distribution
were presented for 1.8 and 1.96 TeV energies at the Tevatron in \cite{NKASV} and for 14 TeV energy at the LHC in Ref. \cite{GKS}. With the calculation of two-loop soft-anomalous dimensions \cite{NKconf,NKRG}, the resummation was extended to
next-to-next-to-leading-logarithm (NNLL) accuracy in \cite{NKRG}. Exact NLO results and approximate NNLO results from NNLL resummation for the $W$-boson $p_T$ distribution were presented at 1.96 TeV energy at the Tevatron and at 7, 8, and 14 TeV energies at the LHC in \cite{NKRG} (see also \cite{NKRGconf}).

In addition to the moment-space pQCD resummation work described above, related theoretical and numerical studies for electroweak-boson production using resummation in Soft-Collinear Effective Theory (SCET) have appeared in \cite{BLS,BBM,BLS2,BBLM}.

In this paper we present the first results for $Z$ production at approximate NNLO using the pQCD moment-space NNLL resummation formalism, and we show the $p_T$ distribution of the $Z$ boson at 1.96 TeV Tevatron energy and at 7, 8, 13, and 14 TeV LHC energies. We also present corresponding results for the $W$-boson $p_T$ distribution, thus extending and updating our previously published results on $W$ production.
We find in general that the approximate NNLO distributions are reliably
estimated at fixed order $\alpha_s^3$ over the experimentally accessible phase
space with $p_T\ge20$ GeV.

In the next section we give some details of the analytical calculation. Section 3 presents numerical results at Tevatron and LHC energies for the $Z$-boson
$p_T$ distribution. Section 4 has the corresponding results for the $W$-boson $p_T$ distribution. We conclude in Section 5.

\mysection{Analytical results}

\begin{figure}
\begin{center}
\includegraphics[width=10cm]{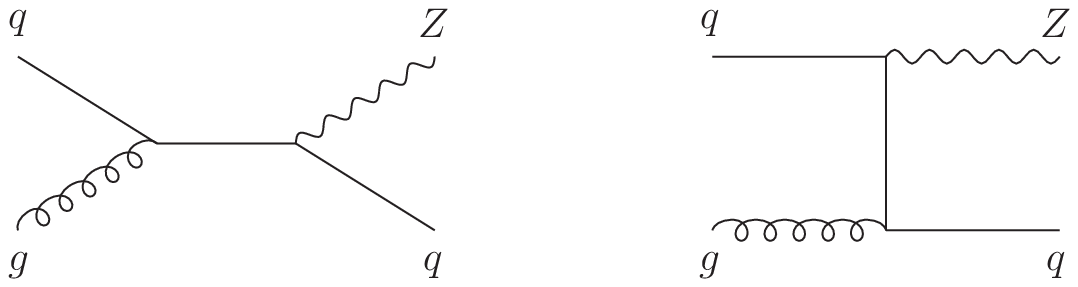}
\includegraphics[width=10cm]{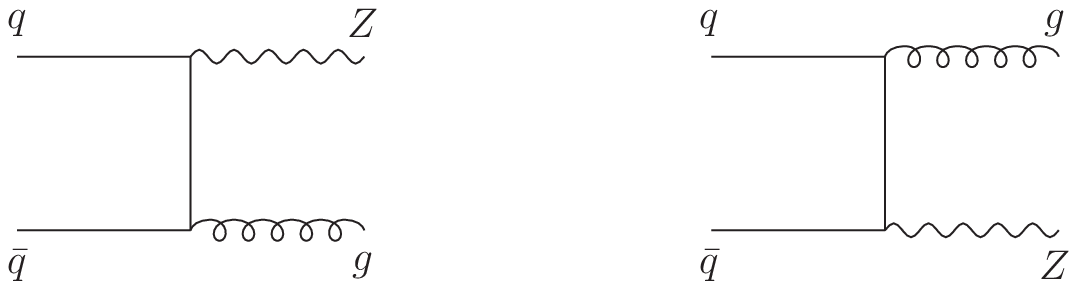}
\caption{LO diagrams for the processes $qg \rightarrow Zq$ (top two graphs) and 
$q{\bar q} \rightarrow Zg$ (bottom two graphs).}
\label{LOdiag}
\end{center}
\end{figure}

We start with the LO contributions to electroweak boson production at
large $p_T$ with a single hard parton in the final state.
The two contributing sub-processes for $Z$ production are (see Fig. \ref{LOdiag})  
$$q(p_a) + g(p_b) \longrightarrow Z(Q) + q(p_c)$$
and
$$q(p_a) + {\bar q}(p_b) \longrightarrow Z(Q) + g(p_c) $$
and similarly for $W$ production.
We define the kinematic variables $s=(p_a+p_b)^2$, $t=(p_a-Q)^2$,
$u=(p_b-Q)^2$, and  note that $p_T^2=tu/s$.
We also define the threshold variable $s_4=s+t+u-Q^2$.
As we approach partonic threshold, where there is no available energy for
additional radiation, $s_4 \rightarrow 0$.

The LO differential cross section for the $qg \rightarrow Zq$ process is
\beq
E_Q \frac{d\sigma^B_{qg\rightarrow Zq}}{d^3Q}
=F^B_{qg \rightarrow Zq} \, \delta(s_4) \, ,
\eeq
with
\beq
F^B_{qg \rightarrow Zq} = \frac{\alpha \,
\alpha_s(\mu_R^2)C_F}{s(N_c^2-1)}
A^{qg} \, |L^Z_{qg}|^2 \, ,
\eeq
\beq
A^{qg} = - \left(\frac{s}{t}+\frac{t}{s}+\frac{2m_Z^2u}{st}\right) \, ,
\eeq
\beq
|L^Z_{dg}|^2=\frac{4\sin^4\theta_W-6\sin^2\theta_W+\frac{9}{2}}{18\sin^2\theta_W\cos^2\theta_W}
\eeq
for $q=d$ (or $s$ or $b$) quark, and
\beq
|L^Z_{ug}|^2=\frac{16\sin^4\theta_W-12\sin^2\theta_W+\frac{9}{2}}{18\sin^2\theta_W\cos^2\theta_W}
\eeq
for $q=u$ (or $c$ or $t$) quark, where $\alpha=e^2/4\pi$, $\alpha_s$ is the strong coupling, $\mu_R$ is the renormalization scale, $m_Z$ is the $Z$-boson mass, $\theta_W$ is the weak mixing angle, and $C_F=(N_c^2-1)/(2N_c)$ with $N_c=3$ the number of colors.

For the process $q{\bar q} \rightarrow Zg$ the LO result is
\beq
E_Q \frac{d\sigma^B_{q {\bar q}\rightarrow Zg}}{d^3Q}
=F^B_{q{\bar q} \rightarrow Zg} \, \delta(s_4) \, ,
\eeq
with
\beq
F^B_{q{\bar q} \rightarrow Zg} =\frac{\alpha \, \alpha_s(\mu_R^2)C_F}{sN_c}
A^{q\bar q}\,  |L^Z_{q\bar q}|^2 \, ,
\eeq
\beq
A^{q\bar q} = \frac{t}{u}+\frac{u}{t}+\frac{2m_Z^2 s}{tu} \, ,
\eeq
with $|L^Z_{d \bar d}|=|L^Z_{dg}|$ and $|L^Z_{u \bar u}|=|L^Z_{ug}|$.

For $W$ production the results are very similar and can be obtained from the above equations with the changes $Z \rightarrow W$, $m_Z \rightarrow m_W$, and $|L^W_{qg}|=|L^W_{q{\bar q'}}|=V_{qq'}/(\sqrt{2}\sin\theta_W)$ with $V_{qq'}$ the CKM matrix elements.

The NLO corrections arise from one-loop parton
processes with a virtual gluon, and real radiative processes with
two partons in the final state. The complete NLO corrections were derived 
in \cite{AR,gpw}. They may be written as
\beq
E_Q\,\frac{d{\sigma}^{(1)}_{f_af_b{\rightarrow}ZX}}{d^3Q}=
\alpha_s^2(\mu_R^2) \left[\delta(s_4) \, B(s,t,u,\mu_R)
+C(s,t,u,s_4,\mu_F) \right]
\label{nlo}
\eeq
where $\mu_F$ is the factorization scale and $X$ denotes additional final-state particles.
The coefficient functions $B$ and $C$ depend on the partons $f_a$, $f_b$ in the
initial state. The coefficient
$B(s,t,u,\mu_R)$ is the sum of virtual corrections and of singular terms
proportional to $\delta(s_4)$ in the real radiative corrections. Coefficient
$C(s,t,u,s_4,\mu_F)$ represents real emission processes away from $s_4=0$.

An important subset of the NLO corrections are those from soft-gluon emission.
We can write the NLO soft and virtual (S+V) corrections for the partonic processes for Z production as
\beq
E_Q\frac{d{\sigma}^{(1) \, {\rm S+V}}_{f_a f_b \rightarrow Z X}}{d^3Q} =
F^B_{f_a f_b \rightarrow Z X}
{\alpha_s(\mu_R^2)\over\pi}\,
\left\{c_3^{f_a f_b} \, \left[\frac{\ln(s_4/p_T^2)}{s_4}\right]_+
+c_2^{f_a f_b} \, \left[\frac{1}{s_4}\right]_+ + c_1^{f_a f_b} \,
\delta(s_4)\right\}
\label{nlosv}
\eeq
and similarly for $W$ production. The coefficients $c_3$ and $c_2$ of the 
plus-distribution terms in Eq. (\ref{nlosv}) can be derived from soft-gluon 
resummation, and their expressions are the same for $Z$ and $W$ production. The leading-logarithm coefficient, $c_3$, depends only on the identities of the incoming partons, but $c_2$ also depends on the details of the partonic process. These coefficients are given by
$c_3^{qg}=C_F+2C_A$ and $c_3^{q \bar q}=4C_F-C_A$;
$c_2^{qg}=-(C_F + C_A) \ln(\mu_F^2/p_T^2)- 3C_F/4$
and
$c_2^{q \bar q}=- 2 C_F \ln(\mu_F^2/p_T^2)
-\beta_0/4$, where $C_A=N_c$ and $\beta_0=(11C_A-2n_f)/3$ is the lowest-order term in the QCD beta function with $n_f=5$ the number of light-quark flavors.
The coefficients of the $\delta(s_4)$ terms are given by
\beq
c_1^{qg}=\frac{1}{2A^{qg}}\left[B_1^{qg}+B_2^{qg} n_f
+C_1^{qg}+C_2^{qg} n_f \right]-\frac{c_3^{qg}}{2}
\ln^2\left(\frac{p_T^2}{Q^2}\right)
+c_2^{qg} \ln\left(\frac{p_T^2}{Q^2}\right)\, ,
\label{c1qg}
\eeq
and
\beq
c_1^{q \bar q}=\frac{1}{2A^{q \bar q}}\left[B_1^{q \bar q}+C_1^{q \bar q}
+(B_2^{q \bar q}+D_{aa}^{(0)}) \, n_f \right]
-\frac{c_3^{q \bar q}}{2} \ln^2\left(\frac{p_T^2}{Q^2}\right)
+c_2^{q \bar q} \ln\left(\frac{p_T^2}{Q^2}\right)\, ,
\label{c1qq}
\eeq
with $B_1$, $B_2$, $C_1$, $C_2$, and $D^{(0)}$
as given in the Appendix of the first paper in Ref. \cite{gpw}
but without the renormalization counterterms
and using $f_A \equiv\ln(A/Q^2)=0$ [note that the terms not multiplying
$A^{qg}$ in Eq. (A4) for $B_1^{qg}$ of Ref. \cite{gpw} should have the
opposite sign than shown in that paper].
Note that Eqs. (\ref{c1qg}) and (\ref{c1qq}) correct the sign of the next-to-last term in Eqs. (4.3) and (4.10) of Ref. \cite{NKRG}. This correction also affects the approximate NNLO numerical results in \cite{NKRG}.

As we will show in the next section, the soft-gluon corrections are numerically important. These corrections can be formally resummed in moment space at NNLL accuracy via the use of two-loop soft anomalous dimensions, the calculation of which involves two-loop diagrams in the eikonal approximation \cite{NKconf,NKRG}. The resummed cross section can then be used as a generator of NNLO approximate corrections \cite{NKRG}.

The NNLO expansion of the resummed cross section  involves logarithms $\ln^k(s_4/p_T^2)$ with $k=0,1,2,3$. At NNLL accuracy the coefficients of all these logarithmic terms can be fully determined, and explicit expressions were already provided in Ref. \cite{NKRG} for $W$ production; the analytical results are identical for $Z$ production and will not be repeated here. In addition, the $\delta(s_4)$ terms involving the factorization and renormalization scales have also been calculated at NNLO \cite{NKASV}. We denote the sum of the exact NLO cross section and the soft-gluon NNLO corrections as ``approximate NNLO.'' 
We will employ the above results, with the noted corrections, to study the $Z$-boson and $W$-boson large-$p_T$ distributions for various collider energies, $\sqrt{S}$, at the LHC and the Tevatron.

\mysection{$Z$-boson production}

In this section we present numerical results for the $p_T$ distribution of the
$Z$ boson in $pp$ collisions at the LHC with $\sqrt{S}=7$, 8, 13, and 14 TeV and in $p{\bar p}$ collisions at the Tevatron with $\sqrt{S}=1.96$ TeV. We use the MSTW2008 \cite{MSTW} parton distribution functions (pdf). We consistently use NLO pdf for the NLO results and NNLO pdf for the approximate NNLO results.

\begin{figure}
\begin{center}
\includegraphics[width=11cm]{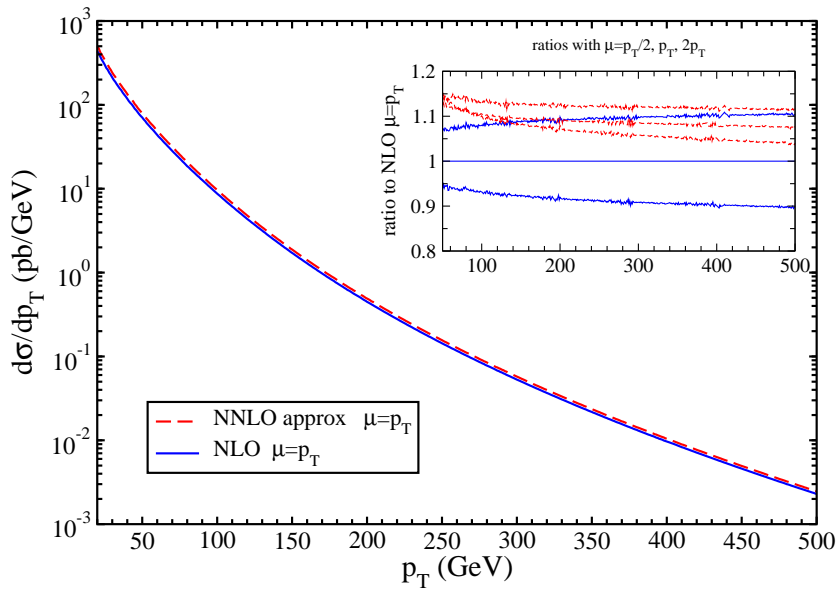}
\caption{$Z$-boson $p_T$ distribution at the LHC at 7 TeV energy.}
\label{Zlhc7}
\end{center}
\end{figure}

We begin with results for $Z$ production at the LHC at 7 TeV energy.
In Fig. \ref{Zlhc7} we plot the $Z$-boson $p_T$ distribution,
$d\sigma/dp_T$, for $p_T$ values up to 500 GeV.
We compare the exact NLO and the approximate NNLO results at 7 TeV LHC energy with central scale $\mu=p_T$. The $p_T$ distribution spans six orders of magnitude in the range of $p_T$ shown in the figure. 
The inset plot shows the ratios of the NLO and approximate NNLO results with different scales, $\mu=p_T/2$, $p_T$, $2p_T$ to the NLO central result with $\mu=p_T$.
The approximate NNLO corrrections provide an increase of the NLO central result, of the order of 10\% for  $\mu=p_T$ (the increase varies from $\sim$13\% at $p_T=50$ GeV to $\sim$8\% at $p_T=500$ GeV).
It is also seen that the scale dependence at approximate NNLO is significantly smaller than at NLO, so there is a decrease in the theoretical uncertainty over all $p_T$ values shown. While at NLO the scale variation is around $\pm 10$\%, at approximate NNLO it is only at most a few percent.

\begin{figure}
\begin{center}
\includegraphics[width=11cm]{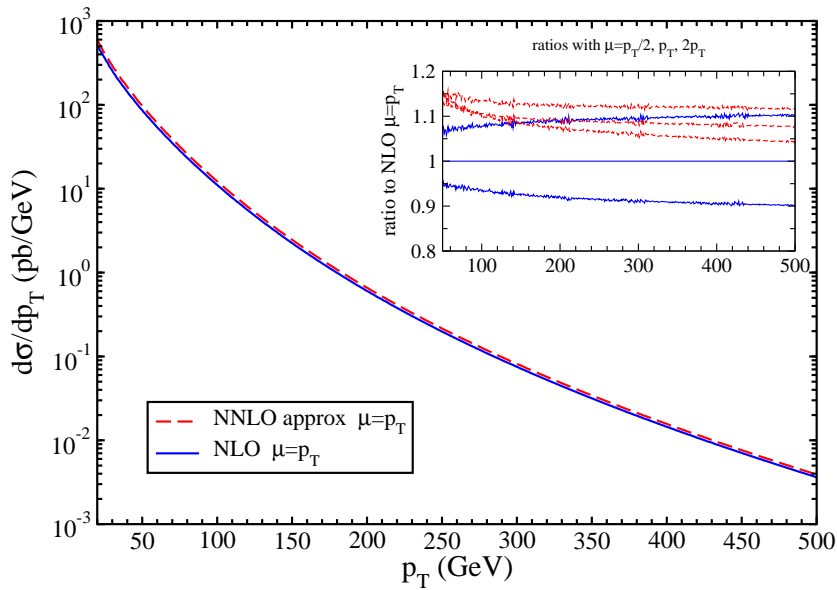}
\caption{$Z$-boson $p_T$ distribution at the LHC at 8 TeV energy.}
\label{Zlhc8}
\end{center}
\end{figure}

We continue with results for $Z$ production at the LHC at 8 TeV energy.
In Fig. \ref{Zlhc8} we again show the $Z$-boson $p_T$ distribution at NLO and approximate NNLO. The inset plot again shows ratios with different scales. Although the overall values for $d\sigma/dp_T$ are higher for this larger energy, the ratios in the inset plots are very similar at 7 and 8 TeV energies.

\begin{figure}
\begin{center}
\includegraphics[width=11cm]{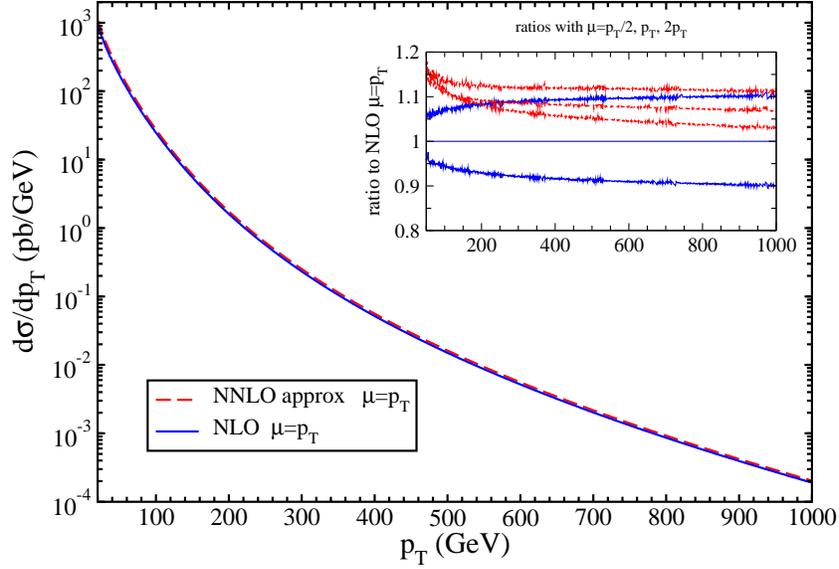}
\caption{$Z$-boson $p_T$ distribution at the LHC at 13 TeV energy.}
\label{Zlhc13}
\end{center}
\end{figure}

In Fig. \ref{Zlhc13} we show the corresponding results at 13 TeV LHC energy and the $p_T$ range shown is up to 1000 GeV. Again, the approximate NNLO corrections enhance the cross section (from $\sim$14\% at $p_T=50$ GeV to $\sim$7\% at $p_T=1000$ GeV, for $\mu=p_T$) while reducing the scale dependence.

\begin{figure}
\begin{center}
\includegraphics[width=11cm]{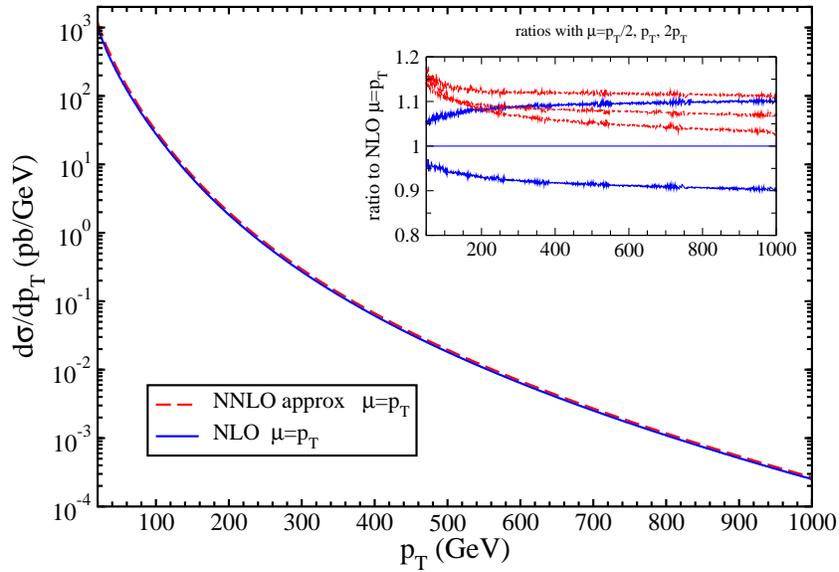}
\caption{$Z$-boson $p_T$ distribution at the LHC at 14 TeV energy.}
\label{Zlhc14}
\end{center}
\end{figure}

Figure \ref{Zlhc14} shows the $Z$-boson $p_T$ distribution for 14 TeV energy at the LHC. The distribution falls over seven orders of magnitude in the $p_T$ range shown. The scale ratios and the overall enhancement from the soft-gluon NNLO corrections at 14 TeV are very similar to those at 13 TeV.

\begin{figure}
\begin{center}
\includegraphics[width=11cm]{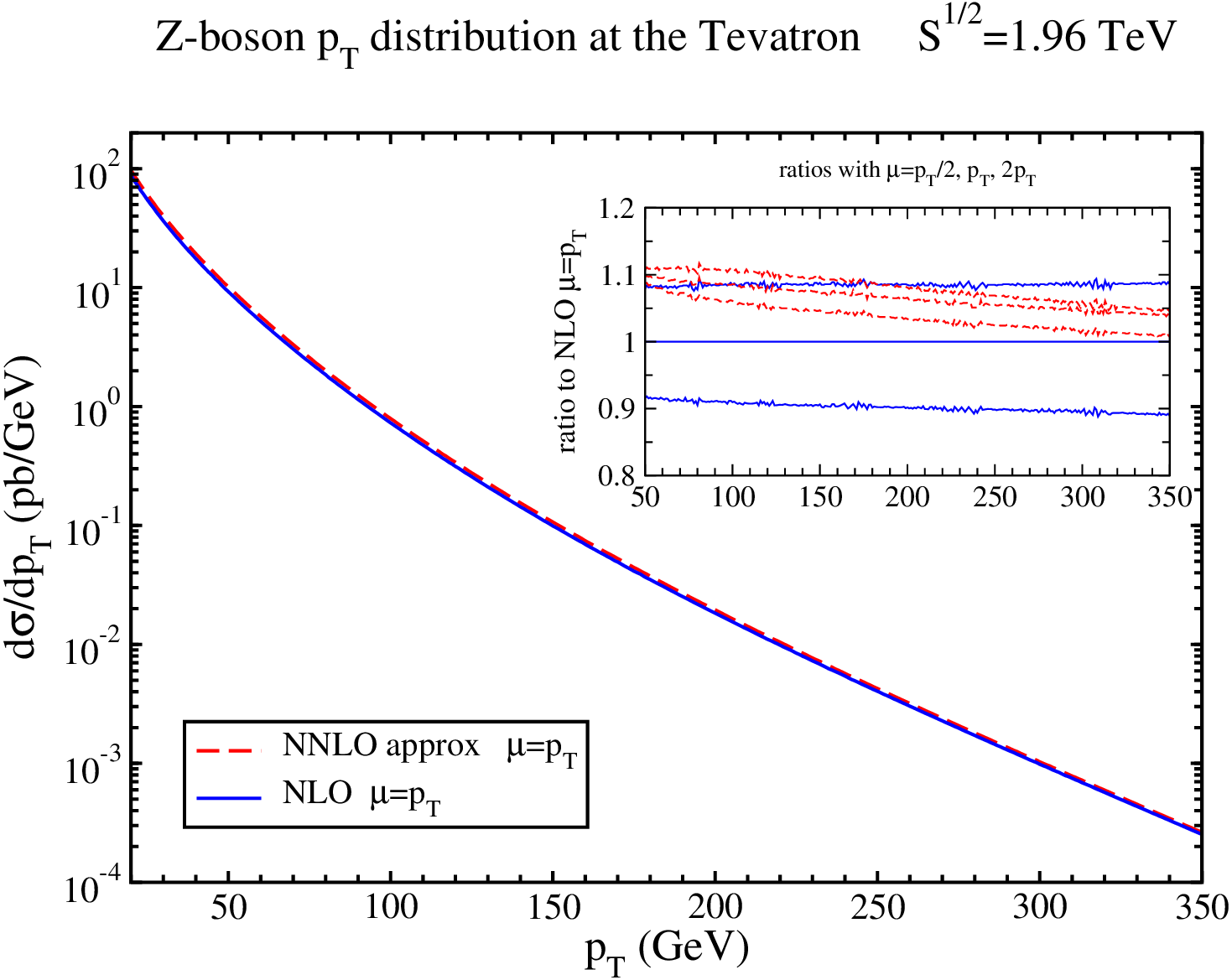}
\caption{$Z$-boson $p_T$ distribution at the Tevatron at 1.96 TeV energy.}
\label{Ztev}
\end{center}
\end{figure}

Finally, in Fig. \ref{Ztev} we show the $Z$-boson $p_T$ distribution at the Tevatron energy of 1.96 TeV for a $p_T$ up to 350 GeV. The inset plot again displays the reduction in scale dependence from around $\pm 10$\% at NLO to a few percent when the approximate NNLO corrections are included. The enhancement from these corrections varies from $\sim$10\% at $p_T=50$ GeV to $\sim$4\% at $p_T=350$ GeV, for $\mu=p_T$.

\begin{figure}
\begin{center}
\includegraphics[width=10cm]{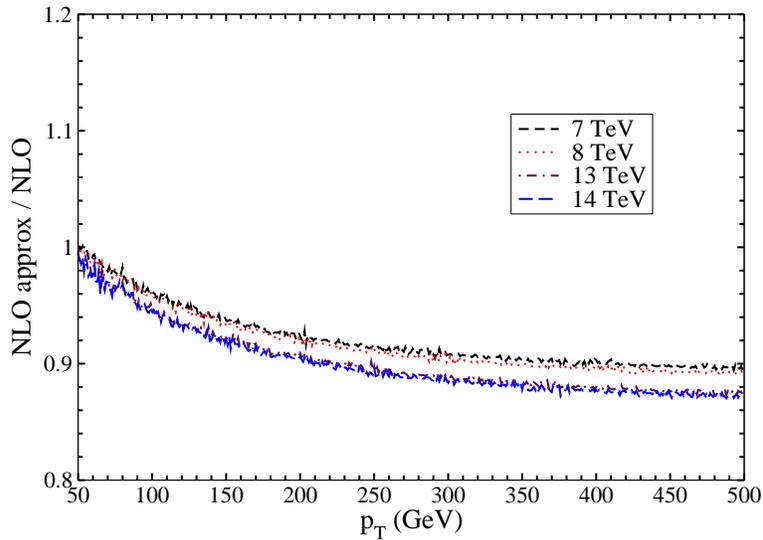}
\caption{Comparison of the approximate and exact NLO $Z$-boson $p_T$ distributions at LHC energies. The 7 and 8 TeV curves (upper two lines) are very similar, as are the 13 and 14 TeV curves (lower two lines).}
\label{Zb1ab1e}
\end{center}
\end{figure}

It is important to check how well the soft-gluon approximation works. In Fig. \ref{Zb1ab1e} we plot the ratio of the NLO approximate (Eq. (\ref{nlosv})) over the NLO exact (Eq. (\ref{nlo})) $Z$-boson $p_T$ distributions at LHC energies. We see that the approximation is very good, with the approximate NLO results being 90\% to 100\% of the full NLO results, depending on the $p_T$ value and the collider energy. At the Tevatron, kinematically closer to threshold, the approximation is even better with the approximate NLO result being 96\% to 100\% of the full NLO value. These results give confidence that the NNLO soft-gluon corrections capture the majority of NNLO contributions and that the approximate NNLO results in Figs. 2-6 are close to the (yet unknown) exact NNLO quantities.

\begin{figure}
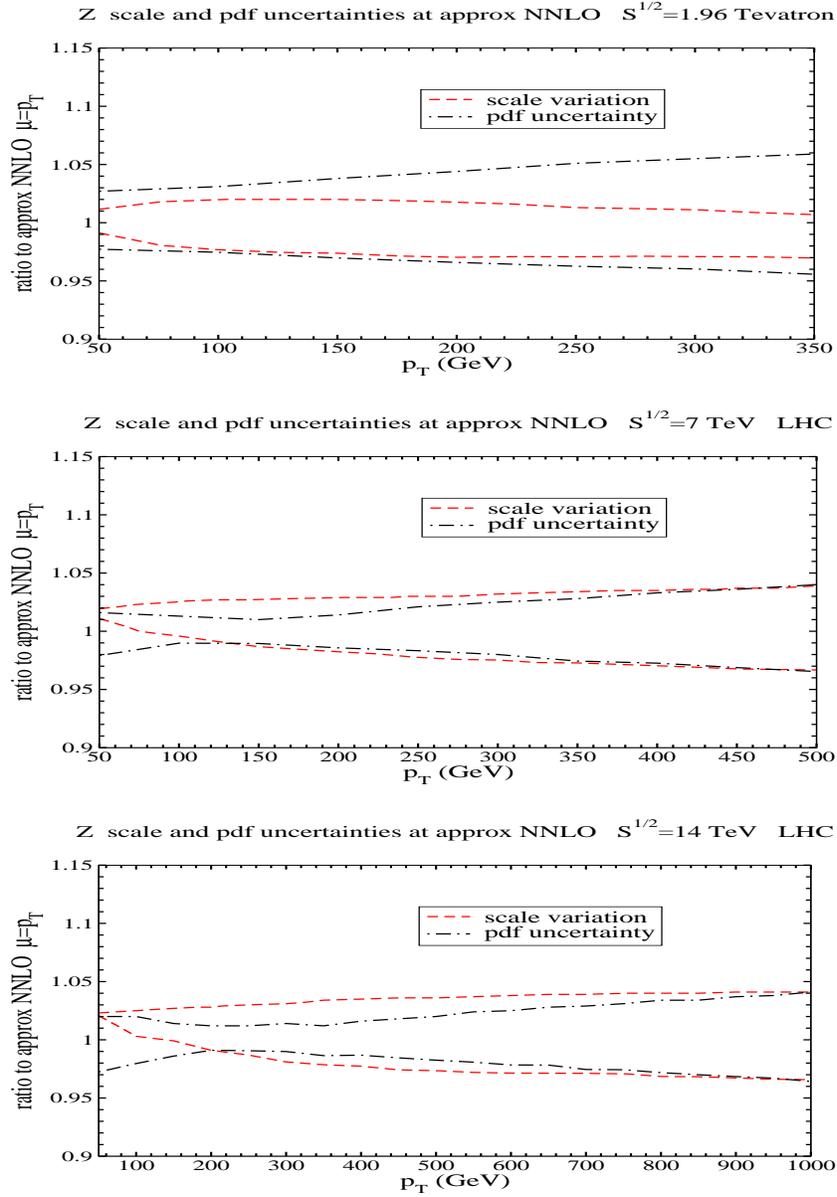

\begin{center}
\includegraphics[width=11cm, height=5cm]{Zpdftevplot.eps}

\vspace{4mm}

\includegraphics[width=11cm, height=5cm]{Zpdf7lhcplot.eps}

\vspace{4mm}

\includegraphics[width=11cm, height=5cm]{Zpdf14lhcplot.eps}
\caption{Comparison of the scale and pdf uncertainties at approximate NNLO for $Z$-boson $p_T$ distributions at Tevatron energy (upper plot) and at LHC energies of 7 TeV (middle plot) and 14 TeV (lower plot). The upper (lower) scale line in each plot is for $\mu=p_T/2$ ($2p_T$).}
\label{Zpdf}
\end{center}
\end{figure}

In addition to scale variation another source of uncertainty comes from the parton distribution function sets used.
In Fig. \ref{Zpdf} we compare  the scale and pdf uncertainties (using MSTW2008 NNLO 90\% CL pdf sets \cite{MSTW}) at approximate NNLO for $Z$-boson $p_T$ distributions at Tevatron and LHC energies. We show ratios relative to the approximate NNLO central set  with $\mu=p_T$. The scale variation is again with $\mu$ ranging from $p_T/2$ to $2p_T$. At the Tevatron the pdf uncertainty is larger than the scale variation for all $p_T$ values shown, especially at higher $p_T$. At LHC energies the pdf uncertainty is somewhat smaller than scale variation for most of the $p_T$ range. Both pdf uncertainties and scale variation are a few percent.

\begin{table}
\begin{center}
\begin{tabular}{|c|c|c|c|c|} \hline
$Z$ boson  & \multicolumn{2}{|c|}{$\sigma$(100 GeV$\le p_T \le p_T^{\rm up})$} & \multicolumn{2}{|c|}{$\sigma$(200 GeV$\le p_T \le p_T^{\rm up})$}
\\ \hline
$\sqrt{S}$ & NLO & NNLO approx & NLO & NNLO approx
\\ \hline
LHC 7 TeV 
&  $277{}^{+24}_{-20}$
&  $305{}^{+8}_{-4}$
&  $20.7{}^{+2.0}_{-1.8}$
&  $22.5{}^{+0.8}_{-0.5}$
\\ \hline
LHC 8 TeV 
&  $360{}^{+29}_{-26}$
&  $395{}^{+11}_{-3}$
&  $28.6{}^{+2.7}_{-2.4}$
&  $31.1{}^{+1.0}_{-0.6}$
\\ \hline
LHC 13 TeV 
&  $863{}^{+66}_{-54}$
&  $951{}^{+24}_{-3}$
&  $84.0{}^{+7.2}_{-6.4}$
&  $91.4{}^{+2.8}_{-1.3}$
\\ \hline
LHC 14 TeV 
&  $979{}^{+72}_{-61}$
&  $1078{}^{+28}_{-2}$
&  $97.6{}^{+8.3}_{-7.3}$
&  $106.2{}^{+3.2}_{-1.5}$
\\ \hline
Tevatron 1.96 TeV 
&  $18.3{}^{+1.6}_{-1.7}$
&  $19.8{}^{+0.4}_{-0.5}$
&  $0.602{}^{+0.051}_{-0.061}$
&  $0.637{}^{+0.009}_{-0.018}$
\\ \hline
\end{tabular}
\caption{NLO and approximate NNLO $Z$-boson cross sections, in pb, integrated over $p_T$ from 100 or 200 GeV to an upper value $p_T^{\rm up}$ which is 350 GeV at the Tevatron, 500 GeV at 7 and 8 TeV LHC energy, and 1000 GeV at 13 and 14 TeV LHC energy.
The indicated uncertainty is from scale variation between $p_T/2$ and $2p_T$.}
\label{pTZtable}
\end{center}
\end{table}

In Table \ref{pTZtable} we present integrated $p_T$ distributions in two
representative $p_T$ bins at the highest experimentally accessible values.
Cross section values in picobarns are given for exact fixed-order NLO and
enhanced approximate NNLO predictions along with scale uncertainties.  
The enhancement from the NNLO soft-gluon corrections over NLO is around 10\%  
for all four LHC energies, and around 8\% at the Tevatron, when integrated over $p_T$ higher than 100 GeV. The scale variation is reduced significantly, by factors of three or four, with the inclusion of these NNLO corrections.
These results would appear to indicate that the predicted event rates are reliably estimated at approximate NNLO.  Deviations from these predicted values would
very likely indicate new physics beyond the Standard Model.

\mysection{$W$-boson production}

In this section we present corresponding numerical results for $W$-boson
production. All results are for the sum of $W^+$ and $W^-$ differential cross sections. 

\begin{figure}
\begin{center}
\includegraphics[width=11cm]{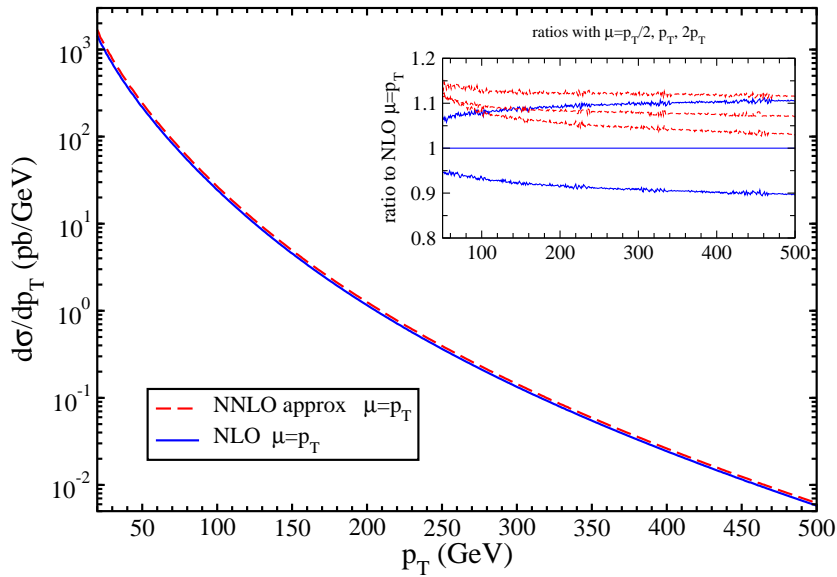}
\caption{$W$-boson $p_T$ distribution at the LHC at 7 TeV energy.}
\label{Wlhc7}
\end{center}
\end{figure}

We begin with results for $W$ production at the LHC at 7 TeV energy.
In Fig. \ref{Wlhc7} we plot the $W$-boson $p_T$ distribution,
$d\sigma/dp_T$, for $p_T$ values up to 500 GeV.
We compare the NLO and the approximate NNLO results with $\mu=p_T$.
The inset plot shows the ratios of the NLO and approximate NNLO results with different scales, $\mu=p_T/2$, $p_T$, $2p_T$ to the NLO result with $\mu=p_T$.
The scale dependence at approximate NNLO is significantly smaller than at NLO, and the approximate NNLO corrrections provide an increase of the NLO central result with a decrease in the theoretical uncertainty over all $p_T$ values shown.
The results are similar to the corresponding ones for $Z$ production shown in the previous section, except that the overall rate is higher for $W$ production.

\begin{figure}
\begin{center}
\includegraphics[width=11cm]{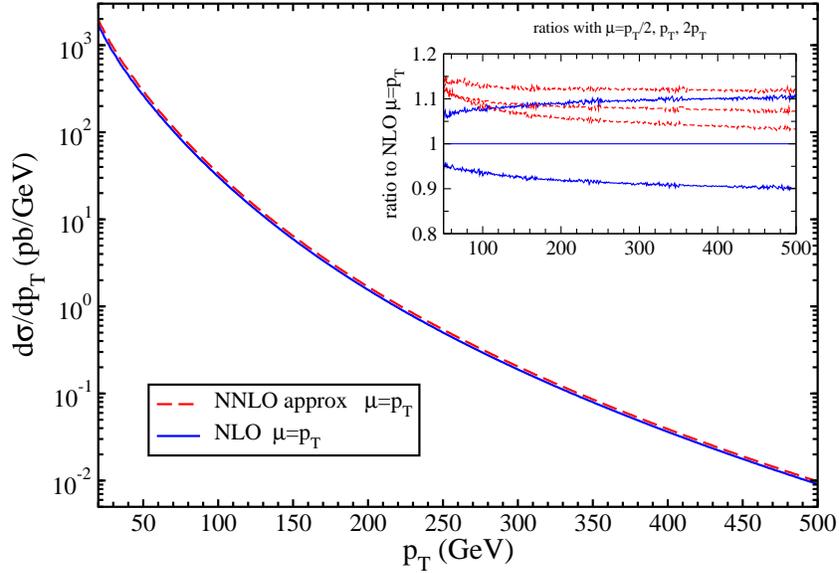}
\caption{$W$-boson $p_T$ distribution at the LHC at 8 TeV energy.}
\label{Wlhc8}
\end{center}
\end{figure}

We continue with results for $W$ production at the LHC at 8 TeV energy.
In Fig. \ref{Wlhc8} we again show the $W$-boson $p_T$ distribution at NLO and approximate NNLO. The inset plot shows ratios of the distributions with different scales. Again, the ratios in the inset plots are very similar at 7 and 8 TeV energies. The approximate NNLO corrections enhance the cross section (from $\sim$11\% at $p_T=50$ GeV to $\sim$7\% at $p_T=500$ GeV, for $\mu=p_T$) while reducing the scale dependence.

\begin{figure}
\begin{center}
\includegraphics[width=11cm]{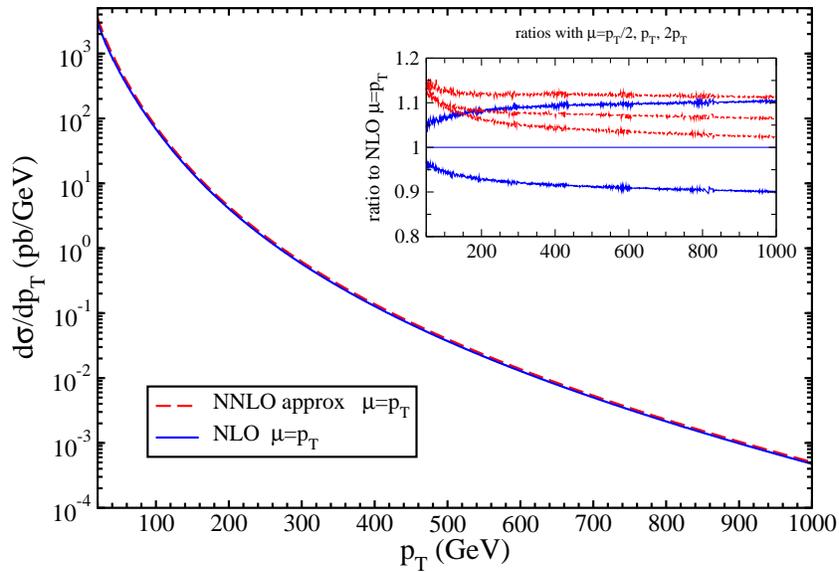}
\caption{$W$-boson $p_T$ distribution at the LHC at 13 TeV energy.}
\label{Wlhc13}
\end{center}
\end{figure}

In Fig. \ref{Wlhc13} we show the corresponding results for the $W$-boson $p_T$ distribution at 13 TeV LHC energy and the $p_T$ range shown is up to 1000 GeV. Again, the approximate NNLO corrections enhance the cross section while significantly reducing the scale dependence.

\begin{figure}
\begin{center}
\includegraphics[width=11cm]{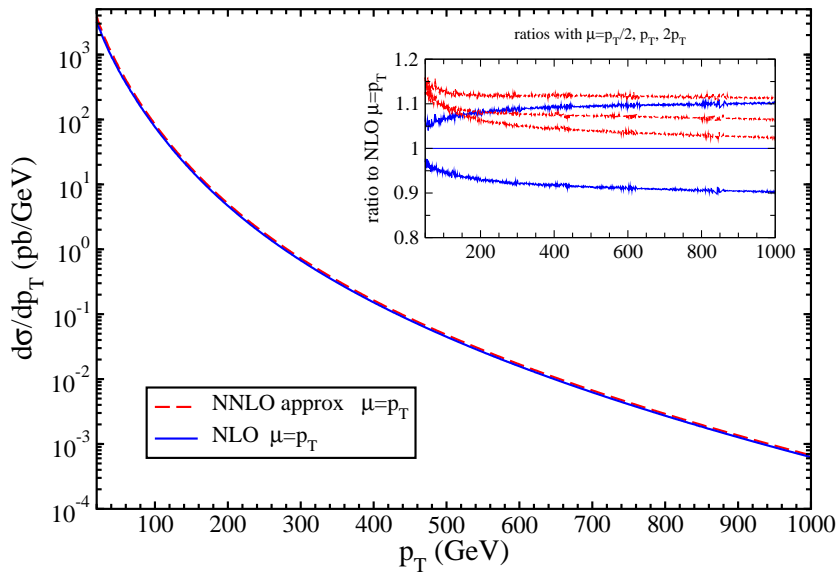}
\caption{$W$-boson $p_T$ distribution at the LHC at 14 TeV energy.}
\label{Wlhc14}
\end{center}
\end{figure}

In Fig. \ref{Wlhc14} we show the $W$-boson $p_T$ distribution for 14 TeV energy at the LHC. The scale ratios at 14 TeV are very similar to those at 13 TeV. The approximate NNLO corrections enhance the cross section (from $\sim$12\% at $p_T=50$ GeV to $\sim$7\% at $p_T=1000$ GeV, for $\mu=p_T$) while reducing the scale dependence. These ratios are also similar to the corresponding ones for the $Z$ boson.

\begin{figure}
\begin{center}
\includegraphics[width=11cm]{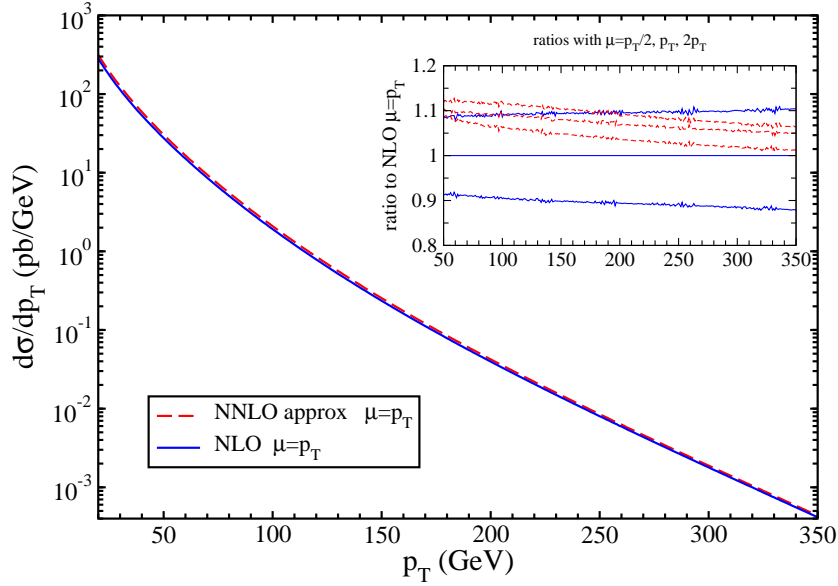}
\caption{$W$-boson $p_T$ distribution at the Tevatron at 1.96 TeV energy.}
\label{Wtev}
\end{center}
\end{figure}

Finally, in Fig. \ref{Wtev} we show the $W$-boson $p_T$ distribution for the Tevatron energy of 1.96 TeV for $p_T$ values up to 350 GeV. The inset plot again displays the enhancement from the approximate NNLO corrections (from $\sim$10\% at $p_T=50$ GeV to $\sim$5\% at $p_T=350$ GeV, for $\mu=p_T$) and the reduction in scale dependence.

\begin{figure}
\begin{center}
\includegraphics[width=11cm, height=5cm]{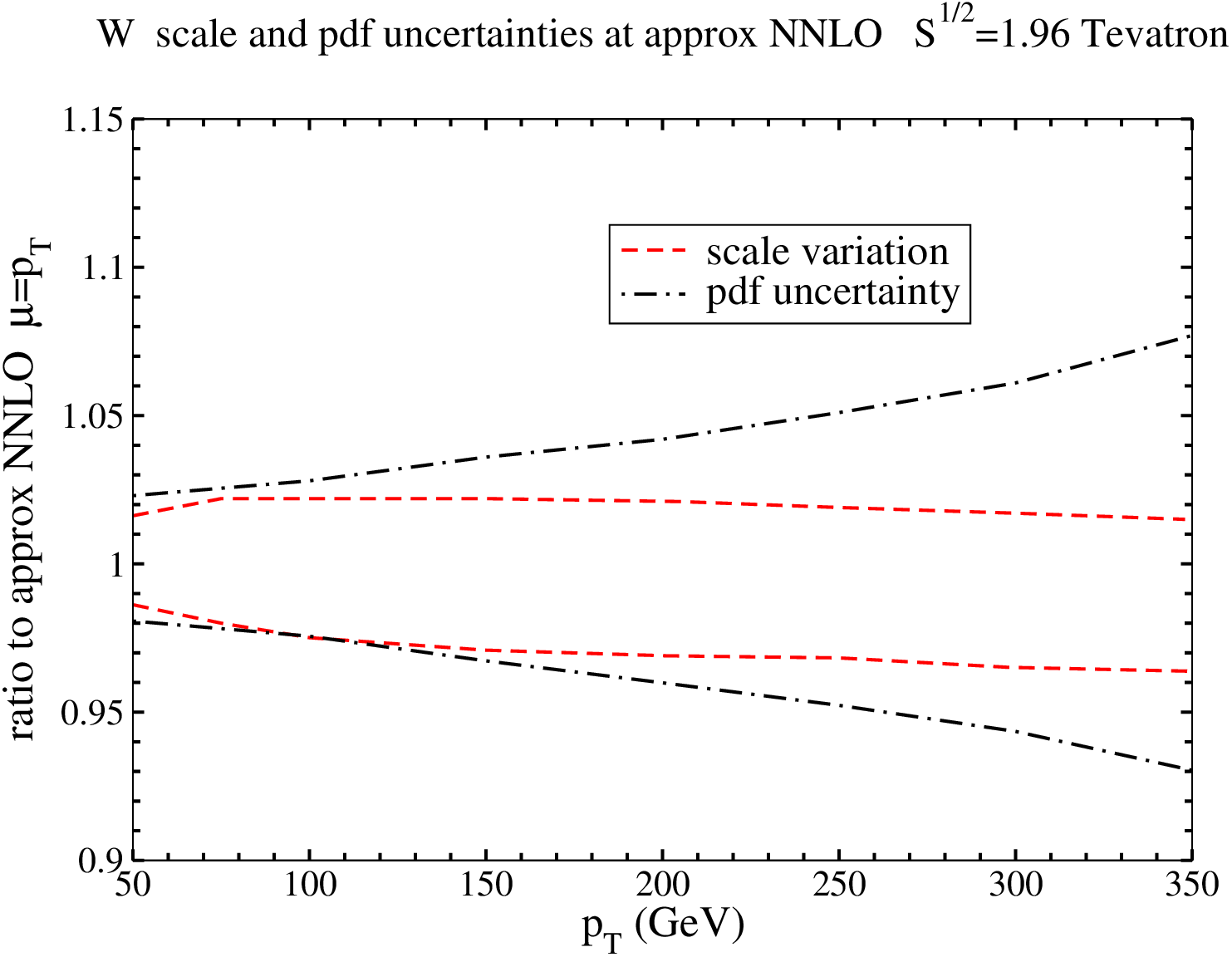}

\vspace{4mm}

\includegraphics[width=11cm, height=5cm]{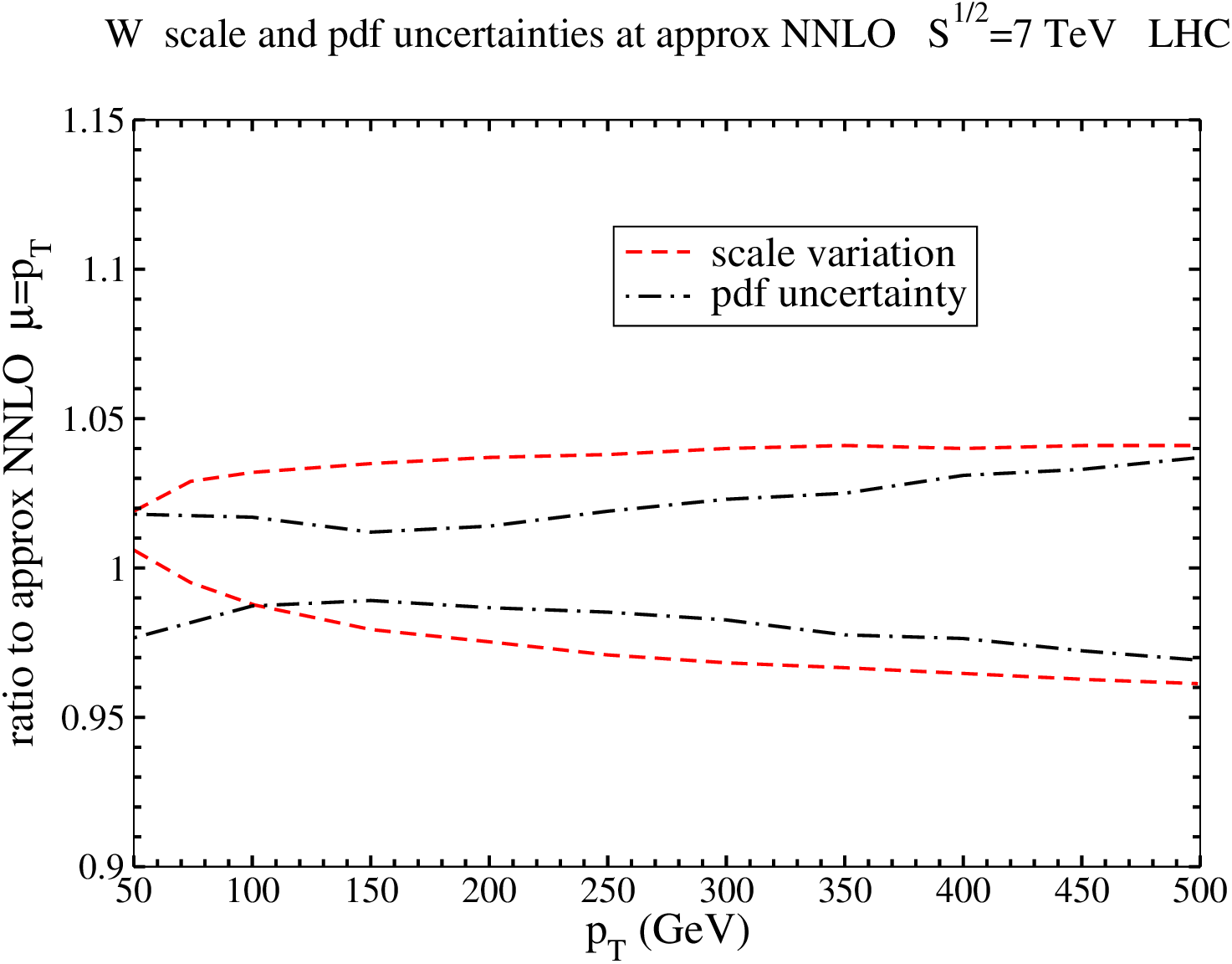}

\vspace{4mm}

\includegraphics[width=11cm, height=5cm]{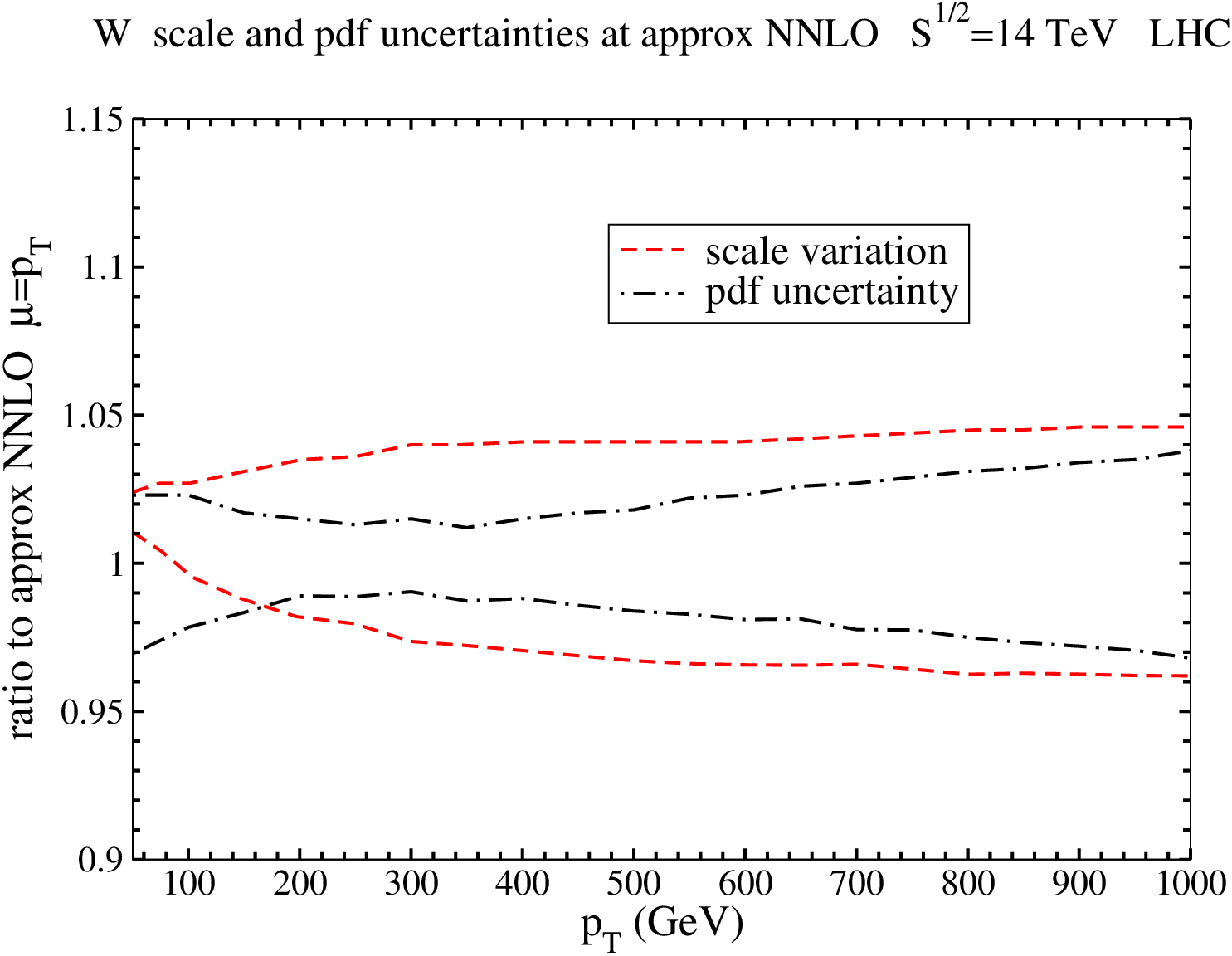}
\caption{Comparison of the scale and pdf uncertainties at approximate NNLO for $W$-boson $p_T$ distributions at Tevatron energy (upper plot) and at LHC energies of 7 TeV (middle plot) and 14 TeV (lower plot). The upper (lower) scale line in each plot is for $\mu=p_T/2$ ($2p_T$).}
\label{Wpdf}
\end{center}
\end{figure}

In Fig. \ref{Wpdf} we compare  the scale and pdf uncertainties at approximate NNLO for the $W$-boson $p_T$ distributions at Tevatron and LHC energies.
Again at the Tevatron the pdf uncertainty is higher than scale variation while at LHC energies the scale variation is larger than the pdf uncertainties for most $p_T$ values. These results are very similar to those for the $Z$ boson in the previous section.

\begin{table}
\begin{center}
\begin{tabular}{|c|c|c|c|c|} \hline
$W$ boson  & \multicolumn{2}{|c|}{$\sigma$(100 GeV$\le p_T \le p_T^{\rm up})$} & \multicolumn{2}{|c|}{$\sigma$(200 GeV$\le p_T \le p_T^{\rm up})$}
\\ \hline
$\sqrt{S}$ & NLO & NNLO approx & NLO & NNLO approx
\\ \hline
LHC 7 TeV 
&  $749{}^{+63}_{-56}$
&  $816{}^{+27}_{-14}$
&  $52.8{}^{+5.1}_{-4.6}$
&  $57.1{}^{+2.2}_{-1.6}$
\\ \hline
LHC 8 TeV 
&  $967{}^{+80}_{-69}$
&  $1054{}^{+35}_{-17}$
&  $72.7{}^{+6.8}_{-6.2}$
&  $78.6{}^{+2.9}_{-2.2}$
\\ \hline
LHC 13 TeV 
&  $2297{}^{+167}_{-143}$
&  $2502{}^{+77}_{-23}$
&  $211{}^{+18}_{-16}$
&  $227{}^{+9}_{-5}$
\\ \hline
LHC 14 TeV 
&  $2599{}^{+188}_{-156}$
&  $2831{}^{+88}_{-23}$
&  $245{}^{+20}_{-19}$
&  $264{}^{+10}_{-6}$
\\ \hline
Tevatron 1.96 TeV 
&  $45.4{}^{+4.2}_{-4.5}$
&  $49.3{}^{+1.1}_{-1.4}$
&  $1.23{}^{+0.12}_{-0.13}$
&  $1.31{}^{+0.03}_{-0.04}$
\\ \hline
\end{tabular}
\caption{NLO and approximate NNLO $W$-boson cross sections, in pb, integrated over $p_T$ from 100 or 200 GeV to an upper value $p_T^{\rm up}$ which is 350 GeV at the Tevatron, 500 GeV at 7 and 8 TeV LHC energy, and 1000 GeV at 13 and 14 TeV LHC energy.
The uncertainty is from scale variation between $p_T/2$ and $2p_T$.}
\label{pT200}
\end{center}
\end{table}

In Table \ref{pT200} we present results for integrated high-$p_T$ $W$-boson
distributions for the same set of parameter values as in Table 1 for $Z$-boson
production.  Once again we note that these rates indicate similar progressive
enhancement from NLO to NNLO with significant reduction in scale uncertainty.  These results indicate that the pQCD predictions for the event rates in these highest $p_T$ bins are reliable for both $Z$ and $W$ production at the Tevatron and the LHC.  The integrated cross sections in the higher bin with $p_T\ge200$ GeV for both $W$ and $Z$ production at the Tevatron and the LHC are compatible with corresponding NNLO$_{\rm sing}$ + NLO results presented in Table 1 of \cite{BLS} and Table 2 of \cite{BBLM} using the SCET formalism. Our results display a somewhat larger enhancement from NNLO soft-gluon corrections and a smaller scale uncertainty than the corresponding ones in \cite{BLS,BBLM}. 

\mysection{Conclusions}

In this paper we have presented theoretical perturbative QCD predictions for
both $Z$-boson and $W$-boson differential cross sections at large $p_T$ at NLO 
and approximate NNLO at the Tevatron and the LHC.
The NNLO soft-gluon corrections increase rates and decrease dependence on
renormalization and factorization scales.  The magnitudes of these effects and
the general trends from LO through approximate NNLO indicate that the
perturbation series is reliably under control.  Since we have shown that at NLO most of the corrections are from soft-gluon emission, it is very likely that the
approximate NNLO corrections provide a reliable estimate of the as yet
uncomputed complete fixed-order $\alpha_s^3$ corrections.  Our results indicate
that it is likely not necessary to add even higher-order soft-gluon effects, beyond the NNLO corrections computed in this paper, in these inclusive $p_T$ distributions at experimentally accessible energies.  These conclusions should also hold true in comparing these QCD predictions to the fiducial cross sections measured experimentally with phase space cuts on the invariant mass, transverse momenta and rapidities of the lepton pairs.

\mysection*{Acknowledgements}
The work of N.K. was supported by the National Science Foundation under
Grant No. PHY 1212472.

\end{document}